\documentclass{article}
\usepackage{slashed}
\newcommand{\bb}{\begin{eqnarray}}
\newcommand{\ee}{\end{eqnarray}}
\begin{document}
\title{Peccei-Quinn-like Symmetries for Nonabelian Axions}
\author{Debashis Chatterjee\\
Bijaygarh Jyotish Ray College, Calcutta 700032\\
P. Mitra\footnote{parthasarathi.mitra@saha.ac.in}\\
Saha Institute of Nuclear Physics, Calcutta 700064, India}
\date{}

\maketitle

\begin{abstract}
Axions were first introduced in connection with chiral
symmetry 
but are now being looked for mainly as
dark matter. In this paper we introduce 
a nonabelian analogue of axions which can also be
potential candidates for dark matter. 
Their nonabelian symmetries, which are generalizations
of the Peccei-Quinn symmetry, are interesting in their own right.
Detailed analysis, using fermion measure and zeta function approaches, 
shows that these symmetries are not anomalous.
\end{abstract}

\section{Introduction}

The chiral symmetry which holds
in classical Dirac theory with massless fermions interacting with gauge
fields is broken by what is called an anomaly \cite{abj}: the transformation 
\bb
\psi\rightarrow e^{i\alpha\gamma_5}\psi,\quad
\bar\psi\rightarrow \bar\psi e^{i\alpha\gamma_5},
\ee
is a symmetry of the kinetic term $\bar\psi[i\slashed{\partial}]\psi$
and also of the interaction term $\bar\psi[i\slashed{A}]\psi$ with the gauge
field $A_\mu$, but the axial current $\bar\psi\gamma_\mu\gamma_5\psi$, which
is classically conserved, is found to violate this
conservation when the fermion triangle diagram is regularized and
evaluated. 
%
The divergence of the axial current 
is finite and proportional to ${\rm tr} F\tilde F$ \cite{bro}.

In general, a classical symmetry may or may not survive quantization. The
simple phase symmetry, whereby $\psi$ is multiplied by a phase factor,
does survive quantization for all masses $m$. 
To see whether a symmetry survives quantization,
the action has to be regularized. If the regularized action still has the
symmetry, the symmetry obviously has no anomaly. If the regularized action
does not possess the symmetry, one tends to think that the symmetry has an
anomaly, but there are different ways of
regularizing fermion field theories and one must check whether a different
regularization can preserve the symmetry.
Familiarity with the chiral anomaly may make one suspect
all classical symmetries involving chiral rotations in any manner
to be anomalous. But whether a symmetry is anomalous or not
has to be checked individually for each symmetry. 

Below we review the example of the Peccei-Quinn symmetry which
occurs in the presence of the hypothetical field called the axion and was 
introduced by these authors. It has 
been shown to survive quantization \cite{pm}. 
We point out that the symmetry can even be made local. 
In section II, a new nonabelian analogue of axions
is introduced: the new Peccei-Quinn symmetry too has no anomaly, as
shown by a measure analysis and a zeta function approach. 

\subsection*{Peccei-Quinn symmetry and the axion}

Chiral symmetry is explicitly broken by the mass term $m\bar\psi\psi$ and also
by quantum effects, {\it i.e.} the anomaly. However, an
artificial chiral symmetry for massive fermions works by letting a new field
${\varphi}$ absorb the chiral transformation. The mass term is replaced by
\cite{pq}
\bb\bar\psi m e^{i{\varphi}\gamma_5}\psi,\ee
which is invariant if the field $\varphi$ transforms under
\bb
\psi\rightarrow e^{i\alpha\gamma_5}\psi,\quad
\bar\psi\rightarrow \bar\psi e^{i\alpha\gamma_5},
\quad\varphi\rightarrow{\varphi}-2\alpha.
\label{0}\ee
This transformation leaves the action invariant provided the new field $\varphi$
is massless. This is the Peccei-Quinn symmetry. The particle \cite{ax}
corresponding to the new field $\varphi$ introduced by them is called the axion,
but it has not been seen in any experiment \cite{ex}. 
It is being studied extensively because it is expected to contribute to
the elusive dark matter. For a discussion of strong CP symmetry in the absence
of axions, one may look at \cite{bcm}.

Careful regularization has been shown to respect the Peccei-Quinn
symmetry, which accordingly is not anomalous but survives quantization 
\cite{pm}.

In spite of this subtlety about the Peccei-Quinn symmetry, the axion can still
be used to remove any $F\tilde F$ term in the action dynamically by coupling
the axion directly to $F\tilde F$ so that its vacuum expectation value
cancels the coefficient of the $F\tilde F$ term.

Observe that
the axial symmetry can be made local by introducing an extra gauge field $B_\mu$
for this purpose:
\bb
\bar\psi[i \slashed{D}+\slashed{B}\gamma_5 -me^{i\varphi\gamma_5}]\psi
+\frac12 F^2(\partial_\mu\varphi+2B_\mu)(\partial^\mu\varphi+2B^\mu),
\ee
where
\bb
B_\mu\rightarrow B_\mu+\partial_\mu\alpha
\ee
under a local chiral transformation.
Here $F$ is a constant of mass dimension such that the axion kinetic term
is $\frac12 F^2\partial_\mu\varphi\partial^\mu\varphi$ and there are
additional kinetic terms of the gauge fields. 
It is to be noted that this provides a formulation of a chiral gauge theory
similar to but different from the Wess-Zumino formulation suggested
in \cite{fs}. The similarity is that in both cases there is an extra degree
of freedom. The difference is that gauge invariance occurs at both the
classical and quantum levels in the present formulation, while in the old
formulation gauge invariance occurs only at the quantum level \cite{new}.


\section{Nonabelian chiral symmetry and nonabelian axions}

The usual chiral symmetry is under a transformation of the fermion in
spinor space. If the fermion is an $SU(N)$ multiplet, there exist nonabelian 
chiral symmetries. The kinetic piece
\bb
\bar\psi i\slashed{\partial}\psi=
\bar\psi_L i\slashed{\partial}\psi_L +\bar\psi_R i\slashed{\partial}\psi_R
\ee
is invariant under the chiral transformations
\bb
\psi_L\rightarrow U_L\psi_L,\quad \psi_R\rightarrow U_R\psi_R,
\label{3}\ee
where $U_L,U_R$ are spacetime independent $SU(N)$ matrices acting on the
two chiral projections of $\psi$.
The gauge interactions will also be invariant under these provided
the matrix $A_\mu$ commutes with $U_L,U_R$.
For instance, the $SU(N)$ could be a flavour group and the colour $SU(3)$ or
the $U(1)$ could be gauged.

The usual mass term $m(\bar\psi_L \psi_R+\bar\psi_R \psi_L)$ is not invariant
under (\ref{3}) unless $U_L=U_R$, in which case of course the transformation is 
not a chiral transformation.
An analogue of the Peccei-Quinn mass term can be introduced: 
$m(\bar\psi_L W\psi_R+\bar\psi_R W^\dagger\psi_L)$.
Here $W$ is a hypothetical $SU(N)$ matrix field analogous to the axion.
Considering that the original 
axion has not been detected, we must be cautious about
such an object.
However, just as the usual axion is expected to be a kind of dark matter,
this nonabelian object too could be relevant as dark matter. 
Note that it is visualized as a new degree of freedom and not as mesonic matter.
The mathematical construction
may in any case be useful for calculations because of the symmetry.
This term is invariant under (\ref{3}) if $W$ transforms as
\bb
W\rightarrow U_LWU^\dagger_R.
\label{4}\ee
As an $SU(N)$ matrix it involves $N^2-1$ parameters which become fields.
The kinetic term for this matrix field has to be of the form
$Tr[\partial_\mu W \partial^\mu W^\dagger]$,
familiar from chiral models. This is invariant under (\ref{4}).
Thus the full action is invariant under the generalized Peccei-Quinn symmetry
\bb
\psi_L\rightarrow U_L\psi_L,\quad \psi_R\rightarrow U_R\psi_R,\quad
W\rightarrow U_LWU^\dagger_R.
\label{5}\ee

The question now is whether this nonabelian symmetry
survives quantization. 
Anomalies arise when regularizations break some symmetries of
classical actions. In the functional integral approach, it is
said that the action has a symmetry which is broken by the measure \cite{f}.

\subsection*{Fermion measure approach}
To formulate the fermion measure, it is customary to expand the fermion
field in eigenfunctions of some operator. To maintain gauge invariance,
the covariant Dirac operator is considered. The eigenvalue equation is
\bb
i\slashed{D}f_n=\lambda_nf_n,
\ee
where the subscript labels the eigenvalue and the eigenfunction.
Under a gauge transformation,
\bb
D_\mu \rightarrow UD_\mu U^{-1},\quad\psi\rightarrow U\psi,
\quad\bar\psi\rightarrow \bar\psi U^{-1},
\ee
so that
\bb
f\rightarrow Uf.
\ee
The field is expanded as
\bb
\psi=\sum_na_nf_n,\quad\bar\psi=\sum_n\bar a_nf^\dagger_n.
\ee
Each $a,\bar a$ is gauge invariant because $\psi$ and
$f$ transform the same way under gauge transformations and $\bar\psi$
and $f^\dagger$ also transform like each other. The gauge invariant measure 
$\prod_n da_nd\bar a_n$ is used for the fermion integration. 
It is well known that
the measure is not chirally invariant: chiral transformations alter $a, \bar a$
and the change of the measure is a Jacobian which can be evaluated after
some regularization and yields the chiral anomaly.
One needs measures for other fields too, but these do not break symmetries.

Given this situation, it would appear that the Peccei-Quinn transformation
would also alter the measure. The above measure would certainly be
altered, but remembering that the requirement of gauge invariance led
to the use of a fermion measure involving the eigenfunctions of the Dirac
operator which contains the gauge field, we can involve the axion field now.
First for the abelian axion, we consider the new expansion
\bb
\psi=e^{-i\varphi\gamma_5/2}\sum_nb_nf_n,\quad
\bar\psi=\sum_n\bar b_nf^\dagger_ne^{-i\varphi\gamma_5/2}.
\ee
Although the fermion field changes under the Peccei-Quinn transformation,
the exponential factor too changes and cancels it, because of (\ref{0}),
leaving $b,\bar b$
invariant. Hence the measure  $\prod_n db_nd\bar b_n$ is invariant
under the transformation. In other words, although the na\"{i}ve fermion
measure is altered by the Peccei-Quinn transformation, there does exist a
fermion measure which is left invariant. This is very similar to what
happens with regularizations. 
It is also similar to the alteration occurring in the fermion measure
in the presence of a twisted fermion mass term \cite{scp}.
It may be added that the measure for $\varphi$
is translation invariant.

For the $SU(N)$ version of axions, the construction of the measure is a bit
complicated. First, note that eigenvalues and eigenfunctions of $i\slashed{D}$
come in pairs:
\bb
i\slashed{D}f_n=\lambda_nf_n,\quad i\slashed{D}\gamma_5f_n=-\lambda_n\gamma_5f_n.
\ee
Hence it is possible to consider expansions in $f_{nL},f_{nR}$, which are
chiral combinations of the $f_n,\gamma_5f_n$, though they are not eigenfunctions
of $i\slashed{D}$. We expand
\bb
\psi_L=\sum_na^L_nf_{nL},&\quad& \psi_R=\sum_na^R_nf_{nR},\nonumber\\
\bar\psi_L=\sum_n\bar a^L_nf^\dagger_{nL},&\quad& \bar\psi_R=\sum_n\bar a^R_nf^\dagger_{nR}.
\ee
Of course, the range of $n$ is implicitly altered here.
The measure $\prod_nda^L_nda^R_nd\bar a^L_nd\bar a^R_n$ is not invariant 
under an $SU(N)$ chiral
transformation (\ref{3}) because $a,\bar a$ have to change unless $U_L=U_R$,
in which case the common {\it vector} transformation may be absorbed in $f$.

However, a new measure can be constructed using the generalized axion field. 
Consider the expansions
\bb
W^\dagger\psi_L=\sum_nb^L_nf_{nL},&\quad& \psi_R=\sum_nb^R_nf_{nR},\nonumber\\
\bar\psi_LW=\sum_n\bar b^L_nf^\dagger_{nL},&\quad& \bar\psi_R=\sum_n\bar b^R_nf^\dagger_{nR}.
\label{b}\ee
As $W$ is invertible, it may also be transferred to the right if desired.
This construction is not unique, but serves the purpose.
Note the asymmetric use of $W$ here. Because of this asymmetry,
the left hand sides of both equations in the first line acquire $U_R$ 
under (\ref{5}) and the left hand sides in the second line acquire 
$U_R^\dagger$, so that the common {\it vector} factor $U_R$ may be absorbed in 
$f$: 
\bb
f_n\rightarrow U_R f_n,
\ee
leaving the $b,\bar b$ invariant. This means that there exists a measure
$\prod_ndb^L_ndb^R_nd\bar b^L_nd\bar b^R_n$ invariant under the $SU(N)$ version
of the Peccei-Quinn transformation, exactly as before.
As regards the measure for $W$, it can be chosen to be $SU(N)$ invariant.
So the measure respects the symmetry and the new nonabelian Peccei-Quinn
symmetry is not anomalous.

\subsection*{Zeta function approach}

Instead of considering a regularized action, one may also look at the
fermion determinant which then has to be regularized.
The most convenient way to do this in this context is the zeta function
regularization \cite{elizalde}. The determinant is that of the Dirac operator, 
which in the simple case of a singlet axion is
\bb
i\slashed{D} - m e^{i{\varphi}\gamma_5}.
\ee
The zeta function regularization works for a hermitian, positive definite
operator, which has to be obtained by constructing the Laplacian,
\bb
\Delta= [-i\slashed{D} - m e^{-i{\varphi}\gamma_5}][i\slashed{D} - m e^{i{\varphi}\gamma_5}].
\ee
Here the gamma matrices have been taken to be antihermitian in euclidean
spacetime.
The determinant of the Dirac operator is defined as the square root of the
determinant of $\Delta$.  The anomaly has been checked in this
framework \cite{reuter}. 

Now one can write
\bb
\Delta= [-i\slashed{D} - m e^{-i{\varphi}\gamma_5}]
e^{i{\varphi}\gamma_5/2}e^{-i{\varphi}\gamma_5/2}[i\slashed{D} - m e^{i{\varphi}\gamma_5}].
\ee
This can be written after some formal manipulations as
\bb
\Delta=
e^{-i{\varphi}\gamma_5/2}
[-i\slashed{D} +\slashed{\partial}\varphi\gamma_5/2- m]
[i\slashed{D} +\slashed{\partial}\varphi\gamma_5/2- m]
e^{i{\varphi}\gamma_5/2}.
\ee
Apart from the initial and final exponential factors,
this depends on the axion field $\varphi$ only through its derivative. 
When the determinant is calculated, those exponential factors cancel. Hence,
its determinant will also involve only
derivatives of this field and will be invariant under translations thereof.
But after the fermion is integrated out, the Peccei-Quinn transformation
is just a constant translation of the axion field, so the 
determinant is invariant under Peccei-Quinn transformations and it has no
anomaly when the product of its eigenvalues
is regularized through the zeta function.

The case of the nonabelian Peccei-Quinn symmetry is more complicated.
Here, the Dirac operator is
\bb
i\slashed{D} -m(WP_R+W^\dagger P_L)
\ee
So one needs the Laplacian
\bb
\Delta=[-i\slashed{D} -m(W^\dagger P_R+WP_L)]
[i\slashed{D} -m(WP_R+W^\dagger P_L)].
\ee
This can be recast as
\bb
\Delta&=&[-i\slashed{D} -m(W^\dagger P_R+WP_L)]\nonumber\\
&&[V^\dagger P_R+YP_L] [VP_R+Y^\dagger P_L]
[i\slashed{D} -m(WP_R+W^\dagger P_L)],
\ee
where $V,Y$ are some $SU(N)$ matrices to be constrained later. This becomes
\bb
\Delta&=&
[V^\dagger P_L+YP_R]
[-i\slashed{D} -i(VP_L+Y^\dagger P_R)\slashed{\partial}(V^\dagger P_R+YP_L)-m]
\nonumber\\ 
&&[i\slashed{D} -i\slashed{\partial}(VP_L+Y^\dagger P_R)(V^\dagger P_L
+YP_R)-m]
[VP_L+Y^\dagger P_R],
\ee
if
\bb
V^\dagger=WY, Y=W^\dagger V^\dagger, VW=Y^\dagger, Y^\dagger W^\dagger=V,
\ee
which can all be satisfied by requiring $V,Y$ to obey the single relation
\bb
VWY=1.
\ee
When the determinant is calculated, the initial and final factors cancel out.
The remaining expression 
\bb
{\rm det}\bigg([-i\slashed{D} -iVP_L\slashed{\partial}V^\dagger 
-iY^\dagger P_R\slashed{\partial}Y-m]
[i\slashed{D} -i\slashed{\partial}VV^\dagger P_L
-i\slashed{\partial}Y^\dagger YP_R -m]\bigg)\nonumber
\ee
involves $V, Y$ only in the combinations
$V\partial V^\dagger, Y^\dagger\partial Y,
\partial VV^\dagger, \partial Y^\dagger Y$.
These are invariant under 
the Peccei-Quinn-like transformations of $W$ whose $U_R$ actions are taken
as left actions on $Y$ and $U_L$ actions as right actions on $V$ because of
the constraint on $VWY$. 
Thus the determinant is invariant under Peccei-Quinn-like transformations.
This persists upon zeta function regularization of the product of its
eigenvalues. Hence
it is seen once again that these symmetries are not anomalous.

As there is no anomaly in the Peccei-Quinn-like symmetries,
gauge fields can again be used to extend these global chiral symmetries to
local ones. For example, for the left handed chiral symmetry, one needs
an $SU(N)$ gauge field matrix $B_\mu$:
\bb
&&\bar\psi[i \slashed{D}+\slashed{B}P_L -m(WP_R+W^\dagger P_L)]\psi\nonumber\\
&+&\frac12 F^2Tr[(\partial_\mu -iB_\mu)W (\partial^\mu W^\dagger+iW^\dagger B^\mu)].
\ee
Here the transformation of the new gauge field is given by
\bb
(\partial_\mu -iB_\mu)\rightarrow U_L(\partial_\mu -i B_\mu)U_L^{-1}.
\ee
Gauge field kinetic terms have to be added.
The right handed chiral symmetry too can be gauged if desired in a similar way.
This reformulation of a chiral gauge theory in a gauge invariant way is again
reminiscent of \cite{fs}.


\section{Conclusion}
Symmetries which appear to be anomalous because they
are not consistent with obvious regularizations may turn out to be
consistent if regularized with care and therefore may be non-anomalous.
The first known case of such a faux anomaly is the Peccei-Quinn symmetry 
which arises in QCD in the
presence of axions, which are now being looked for as dark matter
but have not so far been found. 
Our new $SU(N)$ version of this symmetry,
which holds if $SU(N)$ analogues of axions are introduced,
provides the second example: a fermion measure invariant under this symmetry
has been explicitly constructed and the determinant in the zeta function
approach is also invariant under nonabelian transformations of $W$.
We hope these mathematical observations
will be of as much interest as ordinary axions in the context of dark matter. 
While axions have been expected to be useful in the context of the strong
CP issue, there is no such possibility with the nonabelian analogues discussed
here because they cannot be coupled in a natural way
to the gluon $F\tilde F$ term. However, as
they couple to quarks, they may also transform into other gauge bosons in the
same way as the ordinary axions get feebly converted to photons. They could be
detected for example as decaying
Z bosons just as ordinary axions are sought to be 
caught in the form of photons.
Apart from this, nonabelian axions can be of use in chiral gauge theories
where they are unphysical and get swallowed up while providing mass to gauge
bosons \cite{new}.

\end{document}